\documentclass[11pt,a4paper]{article}

\usepackage{amsfonts}
\usepackage[dvips]{graphicx}
\usepackage{wrapfig}
\usepackage{amsthm}
\usepackage{amsmath}
\usepackage{epsfig}
\usepackage{t1enc}
\usepackage{amssymb}
\usepackage{latexsym}
\usepackage{bm}

\def\p{\partial}
\def\t{\mbox{\hskip 0.4mm tr~\hskip -0.8mm}}
\def\a{\alpha}

\begin{document}

\renewcommand{\figurename}{\small Fig.}

\centerline{\LARGE \bf Mechanics of plates}

\vskip 2mm \centerline{\large Vladim\' ir Balek\footnote{e-mail
address: balek@fmph.uniba.sk}}

\vskip 2mm \centerline{\large \it Department of Theoretical
Physics, Comenius University, Bratislava, Slovakia}

\vskip 1cm

{\small The theory of deformed plates in mechanical equilibrium is
formulated and properties of circular plates lifted at the center
are discussed.}

\vskip 1cm {\bf I. Introduction}

\vskip 4mm The interest in the mechanics of plates dates back to
1809 when the French Academy of Sciences donated an award for the
best work on oscillating plates (Timoshenko 1952). The donation
was iniciated by Napoleon Bonaparte who was impressed by the
demonstration of Chladni patterns at one of the Academy meetings.
Three years later the only contestant, Sophie Germain, submitted a
treatise where she proposed a variation principle corresponding to
a distinctly nonphysical value of Poisson's ratio $\nu = 1$. The
resulting differential equation, when rederived by Lagrange, was
correct, since the correct equation does not contain Poisson's
ratio. Nevertheless the physical basis of the underlying variation
principle remained unclear, which was hardly surprising since the
theory of elasticity did not exist yet. First step towards its
formulation was Poisson's memoir on the theory of plates published
in 1814, shortly after the second, revised version of Germain's
work appeared. Although two years later Sophie Germain was
finally, at third attempt, given the award, the theory of elastic
plates in its present form was formulated by Kirchhoff not earlier
than in 1850. The extension to plastic plates appeared still
later, after the concept of plasticity emerged in the experiments
by Tresca.

\vskip 2mm The theory of elastic plates is nowadays a standard
part of textbooks on the mechanics of deformable bodies; see, for
example, Landau and Lifshitz (1965). The theory of plasticity is
explained in classical books by Hill (1950) and Kachanov (1956),
and mathematics of the {\it deformation} theory of plasticity is
described in detail in the monograph by Temam (1983). Properties
of plastic plates are discussed in Kachanov's book in the chapter
entitled {\it Plane Stress}. Plates are considered in the
framework of the theory of plastic flow, but most results can be
carried over into the deformation theory simply by replacing
velocities by deflections. Other results on plastic plates can be
found in papers published in engineering journals, and still other
results can be obtained by applying dimensional reduction to what
is known from the theory of extended plastic bodies.

\vskip 2mm In this work all pieces are put together and a concise
and self-contained deformation theory of plastic plates is
developed. Throughout the work, the theory of elastic plates is
used as a model. In section II the concepts of elasticity and
plasticity are introduced, in section III constitutive equations
are constructed, in section IV differential equations for plates
in mechanical equilibrium are formulated and boundary conditions
for them are discussed, in section V deformation energy is
computed, in section VI variational principle for deformed plates
is investigated, in section VII two mixed kind of deformations are
discussed and in section VIII some simple examples are presented.


\vskip 4mm {\bf II. Elasticity and plasticity}

\vskip 4mm Elastic deformation is a reversible process: the body
returns to its initial state after the deforming forces have been
lifted. Moreover, the deformation is proportional to the applied
load. This is just a demonstration of the general law stating that
the response is linear in its cause in case the cause is small.
Linear relation between the stress inside the body to which an
external load is applied and the resulting deformation of the body
is called {\it generalized Hooke's law}. Elastic properties of
isotropic materials are characterized by two parameters only, {\it
Young's modulus of elasticity} $E$ (with the physical dimension of
pressure) and {\it Poisson's ratio} $\nu$ (dimensionless).
Consider an homogenous and isotropic beam fastened at one end and
subject to the action of an aligned load at the other end, and
denote relative dilation of the beam in the direction of its axis
by $\epsilon$, relative dilation of the beam in the lateral
direction by $\epsilon'$, and pressure load (force per unit area)
by $\tau$. In the elastic regime it holds
\begin{equation*} \epsilon = \frac 1E \ \tau, \tag{1} \end{equation*}
and
\begin{equation*} \epsilon' = - \frac \nu E \ \tau. \tag{2} \end{equation*}
The first equation is the original Hooke's law, the famous {\it ut
tensio, sic vis} (as the extension, so the force) stated by Robert
Hooke in 1660. The generalized Hooke's law follows from equations
$\thetag 1$ and $\thetag 2$, if one applies them to an
infinitesimal volume of the body.

\vskip 2mm The values of Poisson's ratio are restricted by the
requirement that the deformation energy per unit volume is
positively definite. This requirement implies that Poisson's ratio
must be from the interval $\langle -1, 1/2 \rangle$. If it is 0,
the lateral section of the beam does not change when the beam is
stretched; if it is 1/2, the volume of the beam remains constant;
and if it is from the interval $(0, 1/2)$, the beam contracts in
the lateral direction and its volume increases. For known
materials, Poisson's ratio assumes values typically from 0.3 to
0.4. It equals, for instance, 0.30 for steel, 0.34 for copper and
0.42 for gold.

\vskip 2mm After the pressure load applied to the beam reaches a
certain critical value, called {\it yield limit}, the beam becomes
plastic. In this regime, the deformation is irreversible; after
the external force is removed, some residual dilation of the beam
remains. The linear character of the deformation is lost, too. For
a large class of materials, stretching of the beam proceeds in two
clearly distinguished stages. First the beam dilates while the
force remains more or less unchanged; then the beam continues to
dilate only if the force starts to grow again. This two stages are
called {\it yielding} and {\it hardening}. (The latter term refers
to the fact that if we lift the force and then apply it again,
plasticity takes over only after the force reaches the value from
which it previously jumped to zero. Consequently, a larger force
is needed to start the plastic deformation than for the first
time.) In the course of yielding, Hooke's law in the form $\thetag
1$ is replaced by an even simpler law
\begin{equation*} \tau = \cal E, \tag{3} \end{equation*}
where $\cal E$ is the yield limit. The law is true for $\epsilon_1
< \epsilon < \epsilon_2$, where $\epsilon_1$ and $\epsilon_2$ are
the relative dilations at which yielding and hardening take over
respectively; particularly, $\epsilon_1 = {\cal E}/E$. One
introduces two limit kinds of plasticity: {\it rigid plasticity},
in which the beam does not deform at all until the stress reaches
the yield limit, and {\it perfect plasticity}, in which no
hardening takes place so that the beam stretches to infinity at
the yield limit. Rigid plasticity means $E = \infty$, $\epsilon_1
= 0$, while perfect plasticity means $\epsilon_2 = \infty$. In
fig. 1, the behavior of a deformable beam is represented
schematically by the solid line,
\begin{figure}[ht]
\centerline{\includegraphics[height=5.5cm]{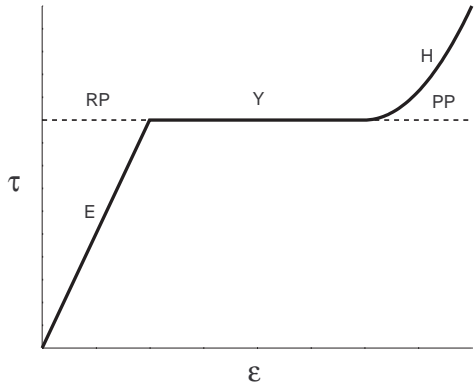}}
\caption{Extension of a beam}
\end{figure}
with 'E' standing for elastic deformation, 'Y' standing for
yielding, and 'H' standing for hardening. The rigid-plastic and
perfect plastic limits are depicted by the dotted-line segments
denoted 'RP' and 'PP' respectively.

\vskip 2mm A realistic description of plasticity is provided by
the {\it incremental theory}, or the {\it theory of plastic flow},
in which the stress inside the body is related to the increment of
the plastic deformation rather that to the plastic deformation
itself. However, if we are interested in the initial stage of
plastic deformation only, we can obtain satisfactory results also
from the {\it deformation theory} in which the stress is related
to the plastic part of the deformation in a similar way as to the
elastic part. In other words, plasticity is regarded as a kind of
a nonlinear elasticity. Particularly, one assigns a deformation
energy to a body, which means that one resigns on the description
of the relaxation of the body to the new equilibrium state after
the load has been lifted. The incremental and the deformation
theory coincide if the body is {\it simply loaded}, which means if
it undergoes, for instance, a dilation or a shear, but not a
dilation followed by a shear. A close connection between the two
theories exists even if the body is arbitrarily loaded, provided
it is rigid-plastic. The deformation in the former theory is then
obtained by integrating the deformation in the latter theory over
time.

\vskip 2mm The simplest theory of plasticity is the deformation
theory of perfectly plastic rigid-plastic bodies. The theory can
be viewed as the rigid limit of the {\it Hencky theory} (the
deformation theory of perfect elastic plasticity), or as the
deformation variant of the {\it Sain-Venant--Levi--von Mises
theory} (the incremental theory of perfect rigid plasticity). Both
theories are discussed in detail in Kachanov (1956). The
deformation theory of perfect rigid plasticity is, despite all its
simplifications, extensively used in engineering. It is the basis
of {\it limit analysis}, evaluation of the loads at which the
plasticity takes over and the corresponding deformations of the
body. In the limit analysis, the assumptions about perfect
plasticity and validity of the deformation theory do not play any
role, and the only remaining simplification, the assumption about
rigid plasticity, provides us with results that can be interpreted
as limit results for real bodies.


\vskip 4mm {\bf III. Constitutive equation}

\vskip 4mm The full system of equations determining the form of a
deformable body in equilibrium consists of {\it constitutive
equation}, {\it equation of balance of forces}, and boundary
conditions for the latter equation. If the body is perfectly
plastic, constitutive equation must be supplemented by {\it
constitutive inequality}. Neither constitutive equation nor
equation of balance of forces are formulated for the plate
directly. One postulates them for the extended body and then
rewrites them into the form valid for the plate. This procedure
can be called {\it dimensional reduction} since one effectively
passes from three to two dimensions. When doing so one assumes
that the plate is thin (its thickness is much smaller than the
characteristic scale of deformation), and that the bending of the
plate is small (the deflection is much smaller than the thickness
of the plate; see Landau and Lifshitz 1965). In the literature
sometimes a different approach is used, based on the original work
by Kirchhoff (see Timoshenko and Woinowsky-Krieger 1959). One
starts from {\it Kirchhoff hypotheses} stating, first, that the
sections that were orthogonal to the mid-plane of the plate at the
beginning remain orthogonal to it after the deforming forces have
been applied, and second, that the mid-plane is neutral, that
means it is bent but neither stretched nor compressed. Both
hypotheses are, however, immediate consequences of the assumptions
cited above.

\vskip 2mm Consider a plate with a constant thickness $h$ whose
mid-plane is placed in the plane $(x,y)$ before the deformation,
and denote the deflection of the mid-plane of the deformed plate
from the plane $(x,y)$ by $w$. The stress inside the plate and the
resulting deformation are described by two symmetric $2 \times 2$
matrices, {\it tensor of moments} $M$ and {\it Hessian matrix}
$W$,
\begin{equation*}
M = \left(\begin{array} {cc}
  M_{xx} & M_{xy}\\
  M_{yx} & M_{yy} \\
  \end{array} \right). \ \
W = \left(\begin{array} {cc}
  W_{xx} & W_{xy}\\
  W_{yx} & W_{yy} \\
  \end{array} \right), \tag{1} \end{equation*}
The components of the Hessian matrix are the second partial
derivatives of the function $w(x,y)$,
$$W = \left(\begin{array} {cc}
  \dfrac {\p^2 w}{\p x^2} & \dfrac {\p^2 w}{\p x \p y}_{}\\
  \dfrac {\p^2 w}{\p x \p y} & \dfrac {\p^2 w}{\p y^2} \\
  \end{array} \right).$$
The matrix is symmetric and its trace equals Laplacian acting on
$w$,
\begin{equation*} \t W = \Delta w. \tag{2} \end{equation*}
The components of the tensor of moments are moments per unit width
induced by the stresses acting parallel to the plate. The moments
$M_{xx}$ and $M_{yy}$ bend the plate in the direction of the axes
$x$ and $y$ respectively, and the moments $M_{xy}$ and $M_{yx}$
twist the plate in the planes $(x,z)$ and $(y,z)$ respectively.
The twisting moments are identical, so that the tensor of moments
is symmetric just like the Hessian matrix.

\vskip 2mm Constitutive equation determines the stress arising
inside the body in terms of deformation. For an elastic body, the
equation reduces to the generalized Hooke's law. Constitutive
equation of a plate is a matrix equation relating the tensor of
moments to the Hessian matrix. It is obtained from the
constitutive equation of an extended body via dimensional
reduction, by which $3 \times 3$ matrices are replaced by $2
\times 2$ ones. If the deformation is elastic and the material of
the plate is isotropic, the constitutive equation reads
\begin{equation*} M = - D [(1 - \nu) W + \nu \t W
I], \tag{3} \end{equation*} where $D = Eh^3/[12(1 - \nu^2)]$. The
constant $D$ is called {\it flexural rigidity}. We can see that
the constitutive equation of the elastic isotropic plate is the
most general relation between two symmetric matrices that is both
linear and isotropic. Since it is obtained by dimensional
reduction of the generalized Hooke's law, it can be called {\it
two-dimensional Hooke's law}.

\vskip 2mm The starting point for the formulation of the
constitutive equation of the plastic body is the {\it yield
criterion}. It is a constraint on the stress inside the body
defining the state of yielding or, if the body is perfectly
plastic, the plastic state itself. The first yield criterion was
proposed in 1868 by Tresca, a French engineer who pioneered the
research of plasticity. Tresca's criterion was later modified by
von Mises. The purpose was to simplify the analysis; however, the
new criterion happened to do even better than the original one
when confronted with experimental data. The yield criterion can be
represented by a surface in the space of main stresses called
{\it yield diagram}. For von Mises' criterion, the yield diagram
is an infinite cylinder with the radius of the base $\sqrt{2/3}
\cal E$, whose axis passes through the origin and is deflected
from all three coordinate axes by the angle $\pi/4$. Von Mises'
criterion can be slightly generalized so that an improved
description of some materials, as marble and sandstone, is
achieved. The generalization was proposed in Yang (1980a); more
specifically, it was the first of the two generalizations proposed
there, describing the effect of hydrostatic stress on yielding. In
the generalized von Mises' criterion materials are characterized,
in addition to the yield limit $\cal E$, by the dimensionless
parameter $\kappa$ assuming values from the interval $\langle 0,
1/2\rangle$. (In the notation of Yang, $\kappa = \alpha (2 -
\alpha)/(1 - \alpha)^2$.) For $\kappa = 0$ the yield diagram is
sphere, for $\kappa$ increasing from 0 to 1/2 it is a gradually
stretching rotational ellipsoid whose axis is deflected by the
angle $\pi/4$ from all coordinate axes, and for $\kappa = 1/2$ it
is von Mises' cylinder. Referring to these shapes we can call the
generalized von Mises' criterion with an arbitrary value of
$\kappa$ {\it ellipsoid criterion}, and the criteria with $\kappa
= 0 $ and $\kappa = 1/2$ {\it spherical criterion} and {\it
cylindric criterion} respectively.

\vskip 2mm If we pass from an extended body to a plate, the yield
criterion undergoes dimensional reduction just like Hooke's law.
As a result, a constraint on the tensor of moments arises.
Particularly, the generalized von Mises' criterion reduces to
\begin{equation*} (1 + \kappa) \t M^2 - \kappa (\t M)^2 = M_0^2. \tag{4}
\end{equation*}
where $M_0 = {\cal E} h^2/4$. Expressed in terms of main moments
(eigenvalues of the matrix $M$), this equation reads
\begin{equation*} M_1^2  + M_2^2 - 2\kappa M_1 M_2 = M_0^2. \tag{5}
\end{equation*}
The yield diagram is now a planar curve that can be obtained as an
intersection of the three-dimensional yield diagram with the
horizontal plane. For $\kappa = 0$ the yield diagram is circle,
while for $\kappa$ increasing from 0 to 1/2 inclusive it is a
gradually stretching ellipse. In fig. 2, three diagrams of this
class are shown,
\begin{figure}[ht]
\centerline{\includegraphics[height=6.5cm]{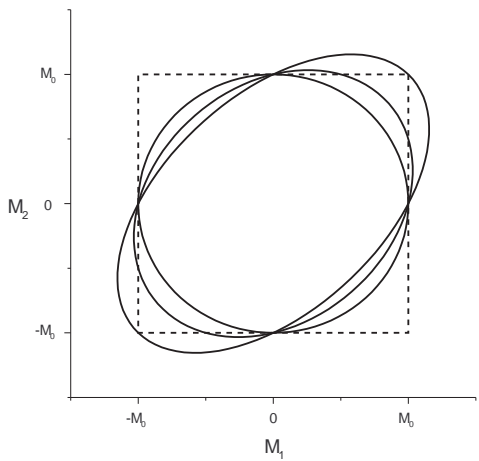}}
\caption{Yield diagrams}
\end{figure}
drawn by the solid line. The form of the diagrams suggests that we
call the yield criterion in question {\it elliptical criterion};
or, if $\kappa = 0$, {\it circular criterion}. One can formulate
other theories of plasticity starting from other yield criteria.
By doing so one does not need to care about the three-dimensional
theory; instead, one can postulate yield criteria that are
two-dimensional from the beginning. Let us mention an alternative
to the elliptical criterion obtained in this way, considered
appropriate for concrete plates. It is the {\it square criterion}
proposed by Johanson (see Mansfield 1957),
\begin{equation*} \text{max} \ (|M_1|, |M_2|) = M_0. \tag{6} \end{equation*}
In fig. 2, this criterion is depicted by the dotted line.

\vskip 2mm If one inserts stresses at the given point of the body
into the yield criterion, one can decide what kind of deformation
occurs there. For a perfectly plastic body in equilibrium, only
two regimes of deformation are possible. If the stresses fit
inside the yield diagram, the deformation is either elastic or
there is no deformation at all depending on whether the body is
elastoplastic or rigid-plastic; if, however, the stresses are
placed on the surface of the yield diagram, the deformation
possibly contains a plastic part. With the stresses outside the
yield diagram there exists no equilibrium. We are talking of an
extended body, but all we have said can be applied to a plate,
too, if the word 'stresses' is replaced by the word 'moments'.
Define the {\it norm} of the tensor of moments $\|M\|$ as the left
hand side of the two-dimensional yield criterion, if written with
$M_0$ on the right hand side. Then the perfectly plastic plate can
be in equilibrium only if the tensor of moments satisfies the
constraint called {\it constitutive inequality},
\begin{equation*} \|M\| \le M_0. \tag{7} \end{equation*}
The plate is deformed elastically or remains flat if $\|M\| <
M_0$, and possibly undergoes plastic deformation if $\|M\| = M_0$.

\vskip 2mm Let us proceed to the constitutive equation for the
plastic plate. Once again, the equation is obtained by dimensional
reduction. We will restrict ourselves to the perfectly plastic
rigid-plastic plate obeying the deformation theory. For such
plate, the Hessian matrix in the regime of plastic deformation is
proportional to minus gradient of the norm $\|M\|$,
\begin{equation*} W = - C \frac {\partial \|M\|}{\partial M}, \ \
C > 0. \tag{8} \end{equation*} In components, the equation reads
$$W_{xx} = - C \frac {\partial \|M\|}{\partial M_{xx}},\ \
W_{xy} = - C \frac {\partial \|M\|}{\partial M_{xy}},\ \ W_{yy} =
- C \frac {\partial \|M\|}{\partial M_{yy}}.$$ For the elliptical
criterion we obtain
\begin{equation*} W = - C_1 [(1 + \kappa) M - \kappa \t M
I], \tag{9} \end{equation*} where $C_1 = C/\|M\| = C/M_0$. Next we
solve for $M$ and fix the constant of proportionality by the yield
criterion. As a result we find
\begin{equation*} M = - {\cal D} \frac {(1 - \kappa)
W + \kappa \t W I} {\sqrt{(1 - \kappa) \t W^2 + \kappa (\t W)^2}},
\tag{10}
\end{equation*} where ${\cal D} = M_0 /\sqrt{1 - \kappa^2}$. This
is the constitutive equation we have sought. After comparing it
with the two-dimensional Hooke's law we conclude that $\cal D$ can
be regarded as 'plastic flexural rigidity' and $\kappa$ as
'plastic Poisson's ratio'.

\vskip 2mm To complete the theory we must specify the behavior of
the plate if the stress is not large enough to produce plastic
deformation. A deformed perfectly plastic rigid-plastic plate
consists, in general, of two parts that are to be treated
separately: the {\it plastic domain}, where $W \ne 0$ and $M$ is
given by the constitutive equation, which means that $\|M\| =
M_0$, and the {\it rigid domain}, where $W = 0$ and $\|M\| \le
M_0$. Obviously, in the rigid domain the plate is flat; or at
least, as we will see, piecewise flat.


\vskip 4mm {\bf IV. Equation of balance of forces}

\vskip 4mm In equilibrium, the forces acting on a column of matter
reaching from one face of the plate to the other must compensate
each other. Obviously, this requirement is the same for elastic
and plastic plates. Suppose the plate is bent by the lateral
pressure load $q$. Equation of balance of forces then reads
\begin{equation*} \nabla \nabla \cdot M = - q, \tag{1}
\end{equation*} where the expression on the left hand side is the
sum of the second derivatives of $M$,
$$\nabla \nabla \cdot M =
\frac {\partial^2 M_{xx}}{\partial x^2} + 2 \frac {\partial^2
M_{xy}}{\partial x \partial y} + \frac {\partial^2
M_{yy}}{\partial y^2}.$$ The double gradient is a symmetric $2
\times 2$ matrix just like the tensor of moments. In general, we
can define the scalar product of matrices as the trace of their
matrix product. Thus, for symmetric matrices we have
$$A \cdot B = \t (AB) = A_{xx} B_{xx} + 2 A_{xy} B_{xy} + A_{yy} B_{yy},$$
which immediately yields the expression for $\nabla \nabla \cdot
M$ above.

\vskip 2mm By combining the equation of balance of forces with the
constitutive equation, we arrive at the differential equation for
deflection only. Consider first an elastic plate. If the material
of the plate is not only isotropic but also homogeneous, $D$ and
$\nu = const$, we find
\begin{equation*} \Delta \t W \equiv \Delta \Delta w = \frac qD. \tag{2}
\end{equation*}
In such a way, the deflection of a homogenous and isotropic
elastic plate in equilibrium obeys an inhomogeneous biharmonic
equation, called {\it Lagrange equation}, with the source
proportional to the pressure load. Note that, as mentioned in the
introduction, the equation does not contain Poisson's ratio.

\vskip 2mm Since the deformation of the plate is governed by a
differential equation of fourth order, we have to impose two
boundary conditions on it. Let us introduce three basic sets of
boundary conditions, corresponding to {\it clamped}, {\it simply
supported} and {\it free} plate. Denote the edge of the mid-plane
of the relaxed plate (a closed curve in the $(x,y)$ plane) by $C$.
The clamped plate is fixed steadily at the edge, hence it
satisfies the conditions
\begin{equation*} w = 0 \ \text{on} \ C, \tag{3} \end{equation*}
and
\begin{equation*} \frac {\partial w}{\partial n} = 0\ \text{on}
\ C, \tag{4} \end{equation*} where $\partial/\partial n$ is the
derivative in the direction normal to the curve. The remaining two
sets of boundary conditions contain also components of the
matrices $M$ and $W$. Denote the unit vectors normal and
tangential to the curve by $\bf n$ and $\bf t$ respectively, and
define normal, mixed and tangential components of a symmetric
matrix as $A_{nn} = A \cdot {\bf nn} = A_{xx} n_x^2 + 2 A_{xy} n_x
n_y + A_{yy} n_y^2$, $A_{nt} = A \cdot {\bf nt} = A_{xx} n_x t_x +
A_{xy} (n_x t_y + n_y t_x) + A_{yy} n_z t_y$ and $A_{tt} = A \cdot
{\bf tt} = A_{xx} t_x^2 + 2 A_{xy} t_x t_y + A_{yy} t_y^2$. The
simply supported plate is fixed by a bar that allows it to rotate
freely, and satisfies the condition $\thetag 3$ together with the
condition
\begin{equation*} M_{nn} = 0 \ \text{on} \ C. \tag{5} \end{equation*}
The constraint on the matrix $M$ follows from the very concept of
mechanical equilibrium: the moment of the bar acting on the plate
must be zero, otherwise the bar would make the plate rotate around
the groove. Rewritten in terms of the matrix $W$, the constraint
reads
\begin{equation*} W_{nn} + \nu W_{tt} = 0 \ \text{on} \ C.
\tag{6} \end{equation*} The free plate is not fixed at all, so
that the bending moment $ M_{nn}$ as well as some effective
shearing force $Q_n$ at its edge must be zero. As a result, the
free plate shares the condition $\thetag 5$ with the simply
supported plate, and in addition it obeys the condition
\begin{equation*} Q_n \equiv {\bf n} \nabla \cdot M  - \frac
{\partial M_{nt}}{\partial l} = 0 \ \text{on} \ C, \tag{7}
\end{equation*} where the first term on the left hand side is the
linear combination of the first derivatives of $M$, with the
coefficients equal to the components of the vector $\bf n$,
$${\bf n} \nabla \cdot M = n_x \frac {\partial M_{xx}} {\partial
x} + n_x \frac {\partial M_{xy}} {\partial y} + n_y \frac
{\partial M_{xy}} {\partial x} + n_y \frac {\partial M_{yy}}
{\partial y},$$ and $\partial/\partial l$ is the gradient in the
direction tangential to the curve $C$. (The scalar product ${\bf
n} \nabla \cdot M$ is a sum of four terms instead of three,
because the matrix ${\bf n} \nabla$ is not symmetric.) Passing to
the Hessian matrix, we find
\begin{equation*} \frac {\partial \t W}{\partial n} - (1 - \nu)
\frac {\partial W_{nt}}
{\partial l} = 0 \ \text{on} \ C. \tag{8} \end{equation*} In
conditions $\thetag 6$ and $\thetag 8$ Poisson's ratio, after all,
enters the theory.

\vskip 2mm Let us summarize. An elastic plate in equilibrium
satisfies the differential equation $\thetag 2$ and two boundary
conditions: conditions $\thetag 3$ and $\thetag 4$, if it is
clamped, conditions $\thetag 3$ and $\thetag 6$, if it is simply
supported, and conditions $\thetag 6$ and $\thetag 8$, if it is
free. From the theory of partial differential equations it follows
that the function solving this problem exists and is unique.

\vskip 2mm Now we wish to formulate conditions of equilibrium of
the plastic plate. Consider a perfectly plastic rigid-plastic
plate whose behavior is described by the deformation theory, and
suppose it obeys the elliptical criterion. Suppose furthermore
that the part of the plate we are interested in lays in the
plastic domain. By inserting from the constitutive equation into
the equation of balance of forces we obtain
\begin{equation*} \nabla \nabla \cdot {\cal W} = \frac q{\cal D}, \tag{9}
\end{equation*}
where
\begin{equation*} {\cal W} = \frac {(1 - \kappa) W + \kappa \t W I}
{\sqrt{(1 - \kappa) \t W^2 + \kappa (\t W)^2}}. \tag{10}
\end{equation*} In such a way, the deflection in the plastic
domain is given by an equation of fourth order again, but unlike
in the elastic case this equation is nonlinear. Boundary
conditions remain the same as in the elastic case, provided we
express the 'moment' and the 'force' conditions in terms of the
tensor of moments. After we pass from the tensor of moments to the
Hessian matrix, the 'moment' condition reduces to $\thetag 6$ with
$\nu$ replaced by $\kappa$, but the 'force' condition is more
complicated than $\thetag 8$ due to the presence of the square
root in the constitutive equation.

\vskip 2mm In the limit analysis, the deformation of a
rigid-plastic plate is called {\it collapse}. If we adopt this
terminology, we can say that the differential equation we have
just obtained is valid everywhere only if {\it total collapse}
takes place. In case of partial collapse we must formulate the
conditions of equilibrium in the rigid domain, too. In this
domain, the equation for deflection is just $W = 0$, however the
deflection is constrained indirectly by the conditions imposed on
the tensor of moments. These include the equation of balance of
forces, the constitutive inequality, and the 'moment' and/or the
'force' condition at the edge of the plate, provided the edge is
rigid, or partly rigid, and the plate is simply supported or free.
The tensor of moments in the rigid domain is independent on the
Hessian matrix locally, but it is correlated with it globally, due
to the requirement that it matches smoothly enough the tensor of
moments in the plastic domain. As a result, the conditions on the
tensor of moments listed above determine the size and shape of the
rigid domain, as well as the shape of the plate in this domain.

\vskip 2mm The fact that the matrix $\cal W$ is a homogenous
function of the Hessian matrix of degree zero, and not of degree
one as in the elastic case, forces us to extend the class of
possible deformations. Consider a deformation by which the plate
has a corner along some line $\hat C$, so that the first
derivative of $w$ in the direction normal to $\hat C$ takes a
finite jump at $\hat C$, and the second derivative has a
$\delta$-function type singularity. In the elastic case, such
deformations are of course forbidden because the corner would
produce a term proportional to the double gradient of the
$\delta$-function in the equation governing the deflection. Now,
however, we have an equation that is not singular at the corner.
Formally, we can see this if we take the square root of the
expression proportional to the $[\delta$-function$]^2$ in the
denominator of the matrix $F$, and cancel the $\delta$-functions
in the numerator and the denominator. As a result, functions with
jumps in the first derivatives must be regarded as potential
solutions to the equation in question. The functional space
containing such functions is called {\it space of functions with
bounded Hessian}. In the rigid domain the plate can have corners
like anywhere else, so that the connected rigid domain is either
flat or composed of several flat pieces sewed together.
Particularly, if the relaxed plate is a polygon, the deformed
plate can be, in the sense it is understood here, entirely rigid.
The borderline between a rigid and a plastic domain can be smooth
(the first derivatives of $w$ can be continuous there), but can be
corner-like as well.

\vskip 2mm If the plate contains corners, the tensor of moments
must satisfy two additional boundary conditions. It must hold
\begin{equation*} M = \pm {\cal D} [(1 - \kappa) {\bf n} {\bf n} +
\kappa I] \ \text{on} \ \hat C, \tag{11}
\end{equation*} where the plus and the minus sign correspond to a
'ridge' and a 'canyon' respectively, and
\begin{equation*} \left(\frac {\partial M_{nn}}{\partial n}\right)_1 =
\left(\frac {\partial M_{nn}}{\partial n}\right)_2 \ \text{on} \
\hat C, \tag{12} \end{equation*} where the indices '1' and '2'
refer to the limits from the two sides of the corner. The
conditions follow from the requirement that moments and shear
forces are balanced along the corner. Note that if the plate is
clamped, the corner can appear at its edge, or a part of its edge,
and if this is the case, condition $\thetag 4$ must be replaced by
condition $\thetag {11}$. In the plastic domain, we must insert
for $M$ from the constitutive equation; condition $\thetag {11}$
then forbids jumps in the second derivatives of $w$, while
condition $\thetag {12}$ restricts jumps in the third derivatives
of $w$. The limit matrix in the former condition is the matrix we
arrive at by the formal procedure described above. Note that we
actually obtain a condition that is apparently much weaker than
$\thetag {11}$, namely
\begin{equation*} M_{nn} = \pm {\cal D} \ \text{on} \ \hat C. \tag{13}
\end{equation*}
However, if we take into account that $M$ obeys the yield
criterion in the plastic domain and the constitutive inequality in
the rigid domain, we can prove that this condition is equivalent
to $\thetag {11}$.

\vskip 2mm A closer look at the conditions of equilibrium we have
established reveals an ambiguity that has no analogue in the
elastic case. If some function is the solution to all conditions,
the same function multiplied by an arbitrary positive constant is
the solution, too. This ambiguity is, however, harmless from the
physical point of view. We obtain a unique solution if we take
into account the hardening. (Formally, there will be still
infinitely many solutions, but only the one corresponding to the
maximal deformation in the yielding regime will become reality.)
Since all solutions differing by a multiplicative factor are
physically equivalent, we can impose an arbitrary normalization
condition on the deflection.

\vskip 2mm From the physical considerations it follows that the
theory has another unusual property that is in a sense dual to the
scaling property discussed above. While for some loads there exist
infinitely many solutions, for other loads there exists no
solution at all. Consider a load of the form $q = \lambda Q$,
where $Q$ is a function of coordinates characterizing the
distribution of the load, and $\lambda$ is a nonnegative
dimensionless parameter characterizing the size of the load. For a
given $Q$, if $\lambda$ is too small, the stresses are not able to
deform the plate, while if $\lambda$ is too large, the stresses
cannot saturate the constitutive inequality. In such a way, a
nontrivial state of equilibrium (such that at least some part of
the plate becomes deformed) occurs only for some limit size, or
sizes, of the load $\lambda_{lim}$. Later we will show that for
any distribution of the load there exists just one
$\lambda_{lim}$.


\vskip 4mm {\bf V. Deformation energy}

\vskip 4mm If we wish to formulate the mechanics of a deformable
body via the variation principle, we need an expression for the
deformation energy (the work done by the stress in the course of
deformation). For the plate, the deformation energy per unit area
is
\begin{equation*} \chi = - \int_0^W M(\bar W) \cdot d\bar W.
\tag{1} \end{equation*} We can rewrite this in terms of the
Hessian matrix, if we use the constitutive equation. Consider
first an elastic plate. By inserting $\thetag {III.4}$ into
$\thetag 1$ we obtain
\begin{equation*} \chi = \frac D2 [(1 - \nu) \t W^2 + \nu (\t W)^2]. \tag{2}
\end{equation*}
Rewritten in terms of the eigenvalues of the Hessian matrix, the
equation reads
\begin{equation*} \chi = \frac D2 (W_1^2 + W_2^2 + 2 \nu W_1 W_2). \tag{3}
\end{equation*}
For a rigid-plastic plate obeying the elliptical criterion, we
insert $\thetag {III.10}$ into $\thetag 1$ and find
\begin{equation*} \chi = {\cal D} \sqrt{(1 - \kappa) \t W^2 + \kappa (\t W)^2]},
\tag{4} \end{equation*} or
\begin{equation*} \chi = {\cal D} \sqrt{W_1^2  + W_2^2 + 2\kappa W_1 W_2}.
\tag{5} \end{equation*} For a rigid-plastic plate obeying the
square criterion, the procedure is a bit more delicate. According
to $\thetag {III.8}$, the eigenvalues of the Hessian matrix $W_1$
and $W_2$ are proportional to the derivatives of $\|M\|$ with
respect to the main moments $M_1$ and $M_2$. At the angles of the
square, the derivatives must be understood in the generalized
sense, as consisting of two limit vectors in the horizontal and
vertical direction and the set of vectors pointing between them.
As a result, we find for nonzero $W_1$ and $W_2$ that $M_1$ and
$M_2$ are equal to $\pm M_0$ with the sign opposite to that of the
corresponding element of the pair $(W_1, W_2)$, and the
deformation energy per unit area is
\begin{equation*} \chi = M_0 \ (|W_1| + |W_2|). \tag{6} \end{equation*}

\vskip 2mm Besides the norm $\|.\|$ in the yield criterion we can
introduce the norm $\|.\|_*$ in the deformation energy. For the
elliptical criterion these norms are
\begin{equation*} \|M\| = \sqrt{(1 + \kappa) \t M^2 - \kappa
(\t M)^2} \tag{7} \end{equation*}
and
\begin{equation*} \|W\|_* = \sqrt{(1 - \kappa) \t W^2 + \kappa
(\t W)^2}. \tag{8} \end{equation*} If $\kappa = 0$, both norms
reduce to the standard (Frobenius) matrix norm, while for other
values of $\kappa$ they differ from the standard norm as well as
from each other. However, there still exists a simple relation
between them. The norms are {\it dual} in the sense that the
square of the latter is, up to a multiplicative factor, the dual
conjugate to the square of the former and vice versa. (To see
that, note that for smooth functions of $n \times n$ matrices the
dual conjugation is just the $n$-dimensional Legendre
transformation.) The norm in the square criterion and that in the
resulting expression for deformation energy are dual, too. This is
in agreement with the general concept of duality in the mechanics
of deformable bodies, discussed in detail in Temam (1983).


\vskip 4mm {\bf VI. Variational principle}

\vskip 4mm With the expression for deformation energy at hand we
can proceed to the variational principle. Denote the region
occupied by the mid-plane of the plate before the deformation by
$S$. The full energy of the plate at rest is
\begin{equation*} U = U_d + U_p, \tag{1} \end{equation*}
where $U_d$ is the deformation energy of the plate,
\begin{equation*} U_d = \int_{S} \chi\ dS, \tag{2} \end{equation*}
and $U_p$ is the potential energy of the plate in the field of
external forces,
\begin{equation*} U_p = - \int_{S} qw \ dS. \tag{3} \end{equation*}
In general, the energy includes also a 'force' and a 'moment'
boundary term; they are, however, both absent for plates that are
fixed in one of the three ways discussed in section IV.
Equilibrium of the plate is determined by the variational
principle
\begin{equation*} U \ \text{is minimal for admissible} \ w\text{'s},
\tag{4} \end{equation*} where the class of admissible $w$'s
depends on how the plate is fixed. Particularly, $w$ must satisfy
the conditions $\thetag {III.3}$ and $\thetag {III.4}$, if the
plate is clamped, the former condition only, if the plate is
simply supported, and no condition at all, if the plate is free.
The plate can be fixed not only at the edge but also at separate
points in the bulk, or along the edge that is otherwise free.
Consider a plate that is fixed at a given set of points $X_a$ at
the heights $h_a$. Clearly, the forces supporting the plate can be
regarded as Lagrange multipliers in the constrained variation
principle
\begin{equation*} U_d \ \text{is minimal for admissible} \
w\text{'s satisfying} \ w(X_a) = h_a. \tag{5} \end{equation*}

\vskip 2mm Euler's equation resulting from the variational
principle $\thetag4$ or $\thetag 5$, when written in terms of the
tensor of moments, is the same for the elastic and plastic plate.
This is immediately seen if we notice that, according to the
definition of the deformation energy, the tensor of moments can be
written as
\begin{equation*} M = - \frac {\partial \chi}{\partial W}.
\tag{6} \end{equation*} By using this relation it is
straightforward to show that Euler's equation for both kinds of
plates is just the equation of balance of forces $\thetag {IV.1}$.
Of course, if we express the tensor of moments from the
constitutive equation, the resulting equation will depend on
whether the plate is elastic or plastic, and in the latter case
also on the choice of yield criterion. Moreover, for plastic
plates Euler's equation, in general, does not hold everywhere. It
surely cannot hold in the rigid domain, since if we attempt to
write it down there, we end up with an undefined expression of the
type 0/0. For certain yield criterions the 'non-Eulerian' domain
can be even more extended. For example, for the square criterion
Euler's equation can be written only in the part of the plastic
domain where the main moments are placed at the angles of the
yield diagram, and in the rest of the plate we must get along with
the constraint on the Hessian matrix and the equation of balance
of forces. In what follows, we will restrict ourselves to the
elliptical criterion, for which the 'non-Eulerian' and rigid
domain coincide.

\vskip 2mm To write down Euler's equation does not mean to solve
the variation principle completely. When we compute $\delta U$,
from integration by parts we obtain, in addition to the surface
integral yielding Euler's equation, the boundary term at the edge
of the plate. If there are corners somewhere throughout the plate,
we get also boundary terms at them. Finally, if the plate is
partly rigid, we arrive at a surface integral over the rigid
domain that cannot be treated in a standard way. We will discuss
these items separately, considering plates of various kinds in the
order of growing complexity: first an elastic plate, then a
rigid-plastic plate that is plastic as a whole, and then a
rigid-plastic plate with a finite rigid domain.

\vskip 2mm For an elastic plate, the only question to discuss is
the form of the boundary term at the edge of the plate. The term
can be written as
\begin{equation*} \delta U(C) = \oint_{C} \left( M_{nn} \frac {\partial
\delta w}{\partial n} - Q_n \delta w \right) dl. \tag{7}
\end{equation*} If the plate is clamped, both terms in the
brackets are identically zero and we obtain no boundary condition
in addition to those determining admissible $w$'s. If the plate is
simply supported or free, the requirement that $\delta U(C)$ is
zero leads to an additional constraint, or constraints, on $w$: we
obtain condition $\thetag {III.5}$ in the former case and
conditions $\thetag {III.5}$ and $\thetag {III.7}$ in the latter
case. In this way we recover the full system of conditions of
equilibrium of an elastic plate established in section IV. Denote
the solution to all conditions by $w_0$. The functional $U[w]$ has
an extreme at $w_0$, $\delta U[w_0] = 0$; and since the functional
is convex, the extreme is absolute minimum.

\vskip 2mm If we use the equation of balance of forces to express
the pressure load in the definition of the potential energy,
integrate two times by parts and use the boundary conditions, we
find
\begin{equation*} U_d = - \frac 12 U_p, \tag{8} \end{equation*}
hence
\begin{equation*} U = - U_d = \frac 12 U_p. \tag{9} \end{equation*}
We can see that the two parts of the full energy, when evaluated
at the solution to the variation principle, obey a simple identity
that can be named {\it virial theorem} after the well-known
theorem from the mechanics of material points. Applying this
identity to a plate that is fixed at separate points, we express
the deformation energy in the form
\begin{equation*} U_d = \frac 12 \sum_a P_a h_a, \tag{10} \end{equation*}
where $P_a$ is the force in the $a$th point.

\vskip 2mm Consider now a fully plastic plate obeying the
elliptical criterion. If the plate contains corners, the integral
defining the full deformation energy has to be understood as
\begin{equation*} \int_S \|W\|_* \ dS= \int_{S \backslash \hat C}
\|W\|_* \ dS + \int_{\hat C} \left| \left[ \frac {\partial
w}{\partial n} \right] \right| \ ds, \tag{11} \end{equation*}
where $\hat C$ is the union of all corners and the square brackets
denote the jump in the quantity inside of them. After calculating
the contributions of the corners to $\delta U$ and putting them
equal to zero, we arrive at conditions $\thetag {IV.11}$ and
$\thetag {IV.12}$. As a result, we obtain the full system of
conditions of equilibrium of a plastic plate formulated in section
IV. The variation of the minimized functional, if evaluated at the
solution, is non-negative, $\delta U[w_0] \ge 0$. (It is positive
if the plate has corners.) Since for the plastic plate the
minimized functional is, just as for the elastic plate, convex,
the solution is again absolute minimum. The minimized functional
is convex but not strictly convex, hence the scaling property of
solutions discussed at the end of section IV. The virial theorem
for the plastic plate states
\begin{equation*} U_d = - U_p, \tag{12} \end{equation*}
so that
\begin{equation*} U = 0, \tag{13} \end{equation*}
and for the plate fixed at a given set of points it holds
\begin{equation*} U_d = \sum_a P_a h_a. \tag{14} \end{equation*}
We can use the scaling property of solutions for a fast derivation
of the virial theorem. Since any solution to the conditions of
equilibrium is absolute minimum of $U$, $U$ must be the same for
all solutions; and since $U$ scales in the same way as $w$ if the
scaling constant is positive, the value of $U$ for all solutions
must be zero.

\vskip 2mm Let us now pass to a partly rigid plate assuming again
that the elliptical criterion is valid. Our starting point will be
an inequality that must be true for any physically acceptable
yield criterion. If $M$ is an arbitrary tensor of moments obeying
the constitutive inequality, and $\hat M(W)$ is the tensor of
moments in the plastic domain corresponding to the given Hessian
matrix $W$, it must hold
\begin{equation*} M\cdot W \ge \hat M(W)\cdot W. \tag{15}
\end{equation*}
This inequality, called {\it Drucker's condition}, guarantees
that, when an elastoplastic plate relaxes after the deforming
forces have been removed, the dissipated energy is non-negative.
As can be seen from the expression of the Hessian matrix in the
form $\thetag {III.8}$, the inequality $\thetag {15}$ is
equivalent to the convexity of the yield diagram in the
three-dimensional space $(M_{11}, M_{22}, M_{12})$; and according
to the two-dimensional version of the theorem proven in Yang
(1980b), the sufficient condition for such convexity is the
convexity of the yield diagram in the two-dimensional space $(M_1,
M_2)$ and the symmetry of $\|M\|$ with respect to the exchanges of
$M_1$ and $M_2$. Consequently, the proof of $\thetag {15}$ for the
elliptical criterion consists in the observation that the
criterion obviously has both properties.

\vskip 2mm To describe the rigid domain in the framework of the
variation principle, let us extend the tensor of moments into it.
When doing so, we must realize that the tensor of moments has
different meaning in the two parts of the plate. In the plastic
domain, it is a $2 \times 2$ matrix formed from the Hessian
matrix, $M = \hat M(W)$; in the rigid domain, it is an auxiliary
$2 \times 2$ matrix that does not depend on deflection. Suppose
the tensor of moments satisfies the equation of balance of forces
throughout the rigid domain. Then we can rewrite the contribution
of this domain to $\delta U$ into the form
$$\delta U(S_r) = \int_{S_r} ({\cal D} \|\delta W\|_* + M\cdot \delta W)
\ dS \ + \ \text{boundary terms},$$
where $S_r$ is the rigid part of $S$.
(We obtain this relation in the same way as we have obtained
the virial theorem.) Suppose, in addition, that the tensor
of moments obeys the constitutive inequality,
\begin{equation*} (1 + \kappa) \t M^2 - \kappa (\t M)^2 \le M_0^2, \tag{16}
\end{equation*}
so that $\thetag {15}$ is valid. For the elliptical criterion, the
matrix $\hat M(W)$ is given by the expression on the right hand
side of $\thetag {III.10}$. Using this expression, we can rewrite
$\thetag {15}$ into the form
$$M\cdot W \ge -{\cal D} \|W\|_*,$$
hence
$$M\cdot \delta W \ge - {\cal D} \|\delta W\|_*,$$
and the first term in $\delta U(S_r)$ is non-negative. We can
easily verify that if $\thetag {16}$ does {\it not} hold, the
first term in $\delta U(S_r)$ can be negative, so that the
validity of $\thetag {16}$ is necessary and sufficient condition
of the non-negativeness of this term. The boundary term coming
from the border between the plastic and the rigid domain is zero
if the matrices $M$ of the two domains match smoothly on the
border. The boundary terms coming from the rigid part of the edge
of the plate, as well as from the corners that lay inside the
rigid domain or at the boundary between the rigid and plastic
domain, are zero if the same boundary conditions hold as for a
fully plastic plate. In such a way, we have obtained the complete
theory of plastic plates of section IV starting from the variation
principle. In the rigid domain we can compute the total
deformation energy in a similar way as in the plastic domain, and
we find that the same virial theorem holds for a partly rigid
plate as for a fully plastic one.

\vskip 2mm We conclude the discussion of the variation principle
for the plastic plate by the proof of uniqueness of the limit
load, mentioned at the end of section IV. Normalize the deflection
of the plate by the condition
\begin{equation*} \int_S wQ\ dS = A, \tag{17} \end{equation*}
where $A$ is a constant with the physical dimension of energy.
If $w_0$ is a solution to the variation principle for a given
$\lambda_{lim}$, from the virial theorem it follows
\begin{equation*} \lambda_{lim} = \frac 1A U_d[w_0]. \tag{18} \end{equation*}
Consider a constrained variation principle
\begin{equation*} U_{d} \ \text{is minimal for admissible normalized} \
w\text{'s}. \tag{19} \end{equation*} This principle is equivalent
to the unconstrained variation principle $\thetag 4$ if we
identify the parameter $\lambda$ with the Lagrange multiplier. We
are accustomed that the Lagrange multiplier is determined by the
corresponding constraint; now, however, this is not the case. The
value assumed by the Lagrange multiplier is given by an expression
analogical to $\thetag {18}$,
\begin{equation*} \lambda^* = \frac 1A U_{d}[w^*], \tag{20} \end{equation*}
where $w^*$ is the solution to the constrained variation
principle. Obviously, $\lambda^*$ is the least of
$\lambda_{lim}$'s. Define the {\it admissible size of the load}
$\lambda_{adm}$ as the value of $\lambda$ for which there exists a
tensor of moments obeying the equation of balance of forces
$\thetag {IV.1}$, the constitutive inequality $\thetag {16}$, and
the boundary condition, or conditions, for the given problem.
According to this definition, all $\lambda_{lim}$'s are
admissible. Furthermore, since $\lambda_{adm} Q =
-\nabla\nabla\cdot M$ and $w^*$ is normalized, it holds
$$\lambda_{adm} = - \frac 1A\int_S \nabla\nabla\cdot M\ w^*\ dS = -
\frac 1A\int_{S^*_p} M\cdot W^*\ dS \le $$
$$\le - \frac 1A\int_{S^*_p} \hat M(W^*)\cdot W^*\ dS = \frac 1A
U_{d}[w^*] = \lambda^*,$$
where $S^*_p$ is the plastic part of $S$ under the deformation described
by $w^*$. (When rewriting the expression for $\lambda_{adm}$ we have
assumed that there are no corners in the rigid domain, but the argument
can be easily generalized to include them.)
In such a way, $\lambda_{adm}$ is not greater than $\lambda^*$,
hence $\lambda_{lim}$ is not greater than $\lambda^*$, and
hence there exists only one $\lambda_{lim}$, equal to $\lambda^*$
as well as to the maximal value of $\lambda_{adm}$.


\vskip 4mm {\bf VII. Two mixed kinds of deformation}

\vskip 4mm A straightforward generalization of the theory of
elastic plates is obtained if one considers a plate that is
stretched. If such plate is bent, an additional lateral force
arises due to the action of stretching forces. Suppose the
stresses induced by the stretching of the plate are homogenous and
isotropic. The additional force per unit area is $- N \t W$, where
the constant $N$, called {\it tension}, is proportional to the
stretching forces. Consequently, the equation of balance of forces
reads
\begin{equation*} \nabla \nabla \cdot M  + N \t W = - q. \tag{1}
\end{equation*}
By inserting for $M$ from the two-dimensional Hooke's law we
obtain
\begin{equation*} D\Delta \t W - N \t W = q. \tag{2} \end{equation*}
To write down the corresponding expression for the deformation
energy, we must determine the work done by the stretching forces
in the course of the bending of the plate. The work equals
$$\frac N2 (\nabla w)^2,$$
where the gradient squared is the sum of the first derivatives
squared,
$$(\nabla w)^2 = \left( \frac {\partial
w}{\partial x}\right)^2 + \left( \frac {\partial w}{\partial y}
\right)^2.$$ As a result, the total deformation energy per unit
surface is
\begin{equation*} \chi = \frac D2 \left[(1 - \nu) \t W^2 + \nu
(\t W)^2\right] + \frac N2 (\nabla w)^2. \tag{3} \end{equation*}
For $N = 0$ we return to equations $\thetag {IV.2}$ and $\thetag
{V.2}$ for a plate without tension, while for $D = 0$ we obtain
the theory of an perfectly flexible plate, or a membrane. Note
that the deformation energy of the membrane is approximately
proportional to the increment of the surface of the membrane due
to the bending. Consequently, the theory of the membrane coincides
with the theory of the bubble, if we consider 'bubbles' in the
form of slightly bent planar layers.

\vskip 2mm In a similar way as we have mixed the deformation of
the plate without tension and the deformation of the membrane, we
can mix elastic and perfectly plastic deformations of the plate,
too. The physical object we arrive at is {\it Hencky's plate}, or
the perfectly plastic elastoplastic plate described by the
deformation theory. In perfectly plastic elastoplastic bodies, a
pure elastic deformation takes place if the stresses lay inside
the yield diagram, and a combined elastic and plastic deformation
possibly takes place if the stresses are placed at the surface of
the yield diagram. Furthermore, the plastic part of deformation is
the same as in rigid-plastic bodies. When passing from three to
two dimensions, we find that the theory is more involved than in
the rigid-plastic case since the plasticity does not take over in
the bulk of the plate at once, but it extends throughout the plate
gradually, penetrating from the faces to the mid-plane. We can,
however, interpolate between the strictly elastic behavior of the
weakly deformed plate and the approximately rigid-plastic behavior
of the strongly deformed plate by adopting the two-dimensional
yield criterion, as well as the expression for the Hessian matrix
of the pure plastic deformation, from the theory of rigid
plasticity. The idea is, just as in the formulation of the square
criterion, to use the pattern of the three-dimensional theory
rather than take this theory as a starting point and derive all
the formulas from it. However, while in the formulation of the
square criterion this shift in perspective was just a shortcut to
the exact theory (we would get the same results from the
three-dimensional theory if we postulated an appropriate
three-dimensional criterion), here it is an approximation. For the
elliptical criterion, we arrive in this way at the previous
formulas describing the plastic domain in the case $\kappa = \nu$,
and formulas with fourth roots otherwise. If $\kappa = \nu$, the
equation for deflection reads
\begin{equation*} D_2 W = q, \ \ D_2 W \equiv
\bigg \{ \begin{array} {l}
D\Delta \t W, \ \ \text{if} \ \ \|W\|_* < {\cal D}/D \\
{\cal D}\nabla \nabla \cdot {\cal W}, \ \ \text{if} \ \ \|W\|_*
\ge
{\cal D}/D \\
\end{array}, \tag{4} \end{equation*}
and the deformation energy per unit area is
\begin{equation*} \chi = \bigg \{ \begin{array} {l}
D \|W\|_*^2/2, \ \ \text{if} \ \ \|W\|_* < {\cal D}/D \\
{\cal D} \|W\|_* - {\cal D}^2/(2D), \ \ \text{if} \ \ \|W\|_* \ge
{\cal D}/D \\
\end{array}. \tag{5} \end{equation*}
For ${\cal D} \to \infty$ and $D \to \infty$ we arrive at the
theory of the elastic and rigid-plastic plate respectively. For
finite $D$ as well as $\cal D$, the whole plate is 'Eulerian' and
the sufficient condition for the existence of the solution to the
problem with the pressure load $\lambda Q$ is the {\it safe load
condition} $\lambda < \lambda_{lim}$ (Temam 1983).


\vskip 4mm {\bf VIII. Examples}

\vskip 4mm The simplest problem in the mechanics of plates is to
determine the shape of the circular plate lifted at the center.
Solution to this problem for an elastic simply supported plate
without tension was found as early as in 1829 by Poisson. In what
follows we will consider plates fixed at the radius 1 with the
center lifted to the height 1, and we will put $D = 1$ for the
elastic plate and ${\cal D} = 1$ for the plastic plate. Note that
the theory is valid only if the deflection is much smaller than
the typical scale on which the plate is deformed, therefore if we
want to obtain physically sensible solutions, we must rescale $w$
by a constant much less than 1. From the symmetry of the problem
it follows $w = w(r)$, where $r$ is the radial coordinate, $r =
\sqrt{x^2 + y^2}$. Deformation of the elastic plate without
tension is given by the equation
\begin{equation*} \Delta_r\Delta_r w = P \delta({\bf r}), \tag{1}
\end{equation*}
where $\Delta_r$ is the radial part of the Laplace operator,
$$\Delta_r = \frac {d^2}{d^2r} + \frac 1r \frac d{dr},$$
$\bf r$ is radius vector and $\delta({\bf r})$ is 2-dimensional
$\delta$-function, $\delta({\bf r}) = \delta(x) \delta(y)$. The
equation must be supplemented by boundary conditions at the center
and the edge of the plate. By assumption, at the center it holds
\begin{equation*} w(0) = 1. \tag{2} \end{equation*}
Suppose the plate is simply supported at $r = 1$. Then the
conditions at the edge are
\begin{equation*} w(1) = w''(1) + \nu w'(1) = 0. \tag{3} \end{equation*}
The general solution to $\thetag 1$ is
$$w = A r^2 \log r + B r^2 + C \log r + D.$$
The condition $\thetag 2$ implies $C = 0$, $D = 1$, the first
condition $\thetag 3$ implies $B = -1$ and the second condition
$\thetag 3$ implies $A = 2(1 + \nu)/(3 + \nu)$. As a result we
obtain
\begin{equation*} w = 1 + r^2 \left[\frac {2(1 + \nu)}{3 + \nu}\log  r
- 1 \right]. \tag{4} \end{equation*} Since $\Delta_r w = 4A \log
r$ and $\Delta_r \log r = 2\pi \delta({\bf r})$, the force is $P =
8\pi A$ and the deformation energy is $U_d = P/2 = 4\pi A$. By
inserting the expression for $A$ into $U_d$ we find
\begin{equation*} U_d = \frac {8\pi(1 + \nu)}{3 + \nu}. \tag{5} \end{equation*}
Suppose now that the plate is clamped at $r = 1$. Then we
have, instead of $\thetag 3$,
\begin{equation*} w(1) = w'(1) = 0. \tag{6} \end{equation*}
This coincides with $\thetag 3$ in the limit $\nu \to \infty$,
therefore the expressions for $w$ and $U_d$ for a clamped plate
can be obtained by performing the limit $\nu \to \infty$ in the
corresponding expressions for a simply supported plate. In
particular, we find that the energy of the plate is $8\pi$. In
fig. 3, two simply supported plates are depicted,
\begin{figure}[ht]
\centerline{\includegraphics[height=6cm]{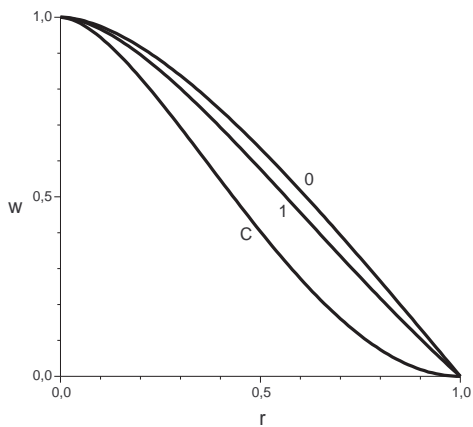}}
\caption{Elastic plates}
\end{figure}
with the values of Poisson's ratio given next to them. (The
nonphysical Sophie Germain's value $\nu = 1$ is used in order to
obtain solutions that are not too close to each other.) In
addition, the curve representing the clamped plate (denoted by
'C') is included into the figure.

\vskip 2mm The one-dimensional version of the problem we are
interested in is the bending of the beam. A well-known result of
the theory of beams is that an elastic beam fixed at the given set
of points assumes the form of {\it cubic spline}, a piecewise
smooth curve used in interpolation problems. The most favored
version of cubic spline is the {\it natural spline} which
corresponds to the beam that is either simply supported at the
endpoints or infinite. The two-dimensional analogue of the natural
spline is the solution describing an infinite elastic plate fixed
at the given set of points, found by Harder and Desmarais (1972)
and known as the {\it thin-plate spline}. One can demonstrate some
features of the thin-plate spline on an infinite plate lifted at
the center and simply supported at its original height by a
ringlike bar that is freely applied to it from above. The
deflection must satisfy, in addition to the condition $\thetag 2$
and the first condition $\thetag 3$, the condition $w'' (\infty) =
0$ (in order that the deformation energy is finite) and the
condition that $w'$ and $w''$ are continuous at $r = 1$ (in order
that the bar does not rotate the plate or produce a corner on it).
By applying these conditions to the general solution written with
different coefficients $A$, $B$, $C$, $D$ in the regions $r \le 1$
and $r > 1$, we obtain
\begin{equation*} w = \bigg \{ \begin{array} {l}
1 + r^2 (\log  r  - 1), \ \ \text{if}\ \ r \le 1\\
- \log  r, \ \ \text{if}\ \ r > 1\\
\end{array}. \tag{7} \end{equation*}
The plate is depicted in fig. 4 by the solid line.
\begin{figure}[ht]
\centerline{\includegraphics[height=6cm]{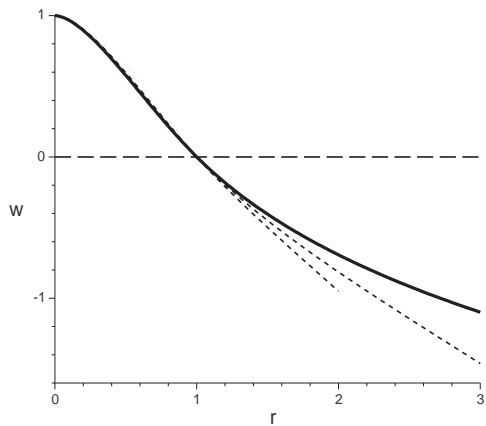}} \caption{An
infinite elastic plate}
\end{figure}
For comparison, some finite free plates with $\nu = 0$ are
presented in the figure, too, drawn by the dotted lines. Their
deflection is
\begin{equation*} w = \bigg \{ \begin{array} {l}
1 + r^2 (p\ \log  r  - 1), \ \ \text{if}\ \ r \le 1\\
- p\ \log  r - q (r^2 - 1), \ \ \text{if}\ \ 1 < r \le R\\
\end{array}, \tag{8} \end{equation*}
where $p = 1/(1 + \Delta)$, $q = 1 - p$, $\Delta = 1/(2R^2)$ and
$R$ is the radius of the plate. The energy of the infinite plate
is $4\pi$ and that of the finite plate is $4\pi p$. Note that
inside the circle at which the plate is fixed, the solution for
the infinite plate coincides with that for the simply supported
plate with the Sophie Germain's value of Poisson's ratio 1.

\vskip 2mm If we include the term describing tension into equation
$\thetag 1$, the solution can be expressed in terms of special
functions. The modified equation reads
\begin{equation*} \Delta_r\Delta_r w - \alpha^2 \Delta_r w =
P \delta({\bf r}), \tag{9} \end{equation*} where $\alpha = \sqrt
N$. To get rid of $\alpha$, replace $\bf r$ by ${\bf u} = \alpha
{\bf r}$ and $P$ by $Q = \alpha^{-2} P$. In this way we obtain
$$\Delta_u \Delta_u w - \Delta_u w = Q \delta({\bf u}),$$ which
is equivalent to
$$\Delta_u \t W - \t W =  Q \delta({\bf u}), \ \ \Delta_u w = W.$$
Equation for $\t W$ is the modified Bessel equation of zeroth
order with the $\delta$-function source. Outside $u = 0$, the
solution is
$$\t W = A I_0(u) + B K_0(u),$$
where $I_0$ and $K_0$ are modified Bessel functions of the first
and second kind, both of zeroth order. The asymptotics of $I_0$
and $K_0$ at $u \sim 0$ are
$$I_0 = 1 + O(u^2), \ \ K_0 = - \log \frac u2 - \gamma +
O(u^2 \log u),$$ where $\gamma = 0.5772...$ is the {\it
Euler-Mascheroni constant}. By using $\Delta_u \log u = 2\pi
\delta({\bf u})$ we find that the equation for $\t W$ is satisfied
at $u = 0$, too, provided $Q = - 2\pi B$. Equation for $w$ yields
$$w = A {\cal I} (u) + B {\cal K} (u) + C \log u + D,$$
where $\cal I$ and $\cal K$ are arbitrarily chosen solutions to
the equations
$$\Delta_u {\cal I} = I_0, \ \ \Delta_u {\cal K} = K_0.$$
The equations have obvious solutions
$${\cal I} = I_0, \ \ {\cal K} = K_0.$$ If we use these solutions
and asymptotics of $I_0$ and $K_0$ given above, we find from the
condition $\thetag 2$ that $C = B$ and $D = 1 - A + B[\log (1/2) +
\gamma]$, so that
\begin{equation*} w = 1 + A F(u) + B G(u), \ \
\bigg \{ \begin{array} {l}
  F = I_0 (u) - 1\\
  G = K_0 (u) + \log \dfrac u2 + \gamma\\
\end{array}. \tag{10} \end{equation*}
The constants $A$ and $B$ must be determined from the conditions
$\thetag 3$. For the plate with $\nu = 0$ we find
\begin{equation*} w = 1 - \frac {qF(\a r) - pG(\a r)}{qF(\a) -
pG(\a)}, \ \ \bigg \{\begin{array} {l}
  p = F''(\a) = I_0(\a) - \dfrac 1\a I_1(\a)\\
  q = G''(\a) = K_0(\a) + \dfrac 1\a K_1(\a) - \dfrac 1{\a^2}\\
 \end{array}. \tag{11}
\end{equation*}
The force is given by the coefficient $B$ appearing in front of
the function $G$ in the expression for $w$, $P = Q \a^2 = - 2\pi B
\a^2$. The deformation energy can be computed again as $U_d = -
U_p/2 = P/2$, hence its value for the plate with $\nu = 0$ is
\begin{equation*} U_d = - \frac {\pi p \a^2}{qF(\a) -
pG(\a)}. \tag{12} \end{equation*} In fig. 5, some plates with $\nu
= 0$ are depicted by the solid line with the values of $\alpha$
given next to them.
\begin{figure}[ht]
\centerline{\includegraphics[height=6cm]{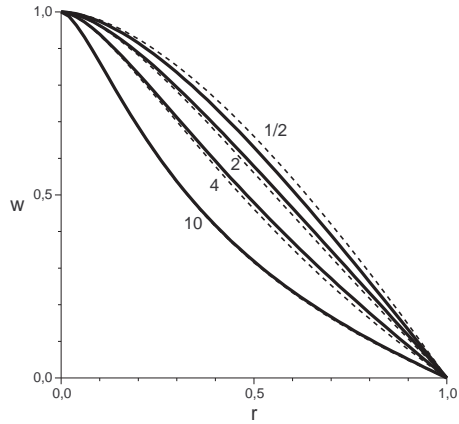}}
\caption{Elastic plates with tension}
\end{figure}
An extended version of thin-plate splines that includes tension
has been proposed by Franke (1985). The base functions entering
these splines are drawn in the figure by the dotted lines. They
are obtained by replacing the condition $w''(1) = 0$ (the second
condition $\thetag 3$) by the nonphysical condition $w''(\infty) =
0$, and are given by the formula
\begin{equation*} w = 1 - \frac {G(\a r)}{G(\a)};
\end{equation*}
see Mit\'a\v s and Mit\'a\v sov\'a (1988).

\vskip 2mm Let us now pass to the plastic plate. If the elliptical
criterion is valid, the deflection of the plate is given by the
equation
\begin{equation*} \nabla \nabla \cdot {\cal W} = P \delta({\bf
r}). \tag{13} \end{equation*} If, moreover, the plate is simply
supported, $w$ obeys the same boundary conditions as before, with
$\nu$ replaced by $\kappa$. Consider $w$ of the form
\begin{equation*} w = 1 - r^p. \tag{14} \end{equation*}
From the definition of $\cal W$ we obtain
$${\cal W} = const + A {\bf n} {\bf n},$$
where ${\bf n} = {\bf r}/r$ and $A = (1 - \kappa)(1 + q)/\tilde
q$, $q = 1 - p$, $\tilde q = \sqrt{q^2 - 2\kappa q + 1}$. This
yields
$$\nabla \nabla \cdot {\cal W} = A \left[ \frac \partial {\partial x}
\left(\frac {n_x}r \right) + \frac \partial {\partial y}
\left(\frac {n_y}r \right)\right] = 2\pi A \delta({\bf r}),$$ so
that the {\it ansatz} $\thetag {14}$ solves equation $\thetag
{13}$ if $P = 2\pi A$. From the second condition $\thetag 3$ with
$\nu$ replaced by $\kappa$ we obtain
\begin{equation*} p = 1 - \kappa, \tag{15} \end{equation*}
and if we insert this into the definition of $A$ and use $U_d = P
= 2\pi A$, we find
\begin{equation*} U_d = 2\pi \sqrt{1 - \kappa^2}. \tag{16} \end{equation*}
For $\kappa = 0$ the plate has the shape of conus, and with
increasing $\kappa$ it bends inside. Note that the conus is
obtained also in the theory with the square criterion, see
Kachanov (1956).

\vskip 2mm Solution $\thetag {15}$ applies also to {\it obliquely}
clamped plates which satisfy, in addition to the condition $w(1) =
0$, the condition $w'(1) = p$ with an arbitrary positive $p$. The
shape of such plates is given by equation $\thetag {15}$
irrespective of their value of $\kappa$. This rises a question as
to what is the shape of the ordinary clamped plate that has $w'(1)
= 0$. To answer that, introduce the function
\begin{equation*} \xi = \lim \limits_{p \to 0^+}(1 - r^p)
\approx \bigg \{\begin{array} {l}
  1\ \text{if}\ r = 0\\
  0\ \text{if}\ 0 < r \le 1 \\
 \end{array}. \tag{17}
\end{equation*}
The explicit expression for $\xi$ is symbolic only, since the
limit is understood in the weak sense, as an operation to be
performed {\it after} the rest of the computation has been
completed. In particular, if we define the integral norm of the
function $w(r)$ as the integral of the Frobenius norm of its
Hessian matrix,
$$\| w \| = \int \sqrt{\t W^2} dS = 2\pi \int \limits_0^1 \sqrt{r^2 w''^2 +
w'^2} dr,$$ we find that the norm of $\xi$ is finite and equals $2
\sqrt{2} \pi$. In fact, we can define $\xi$ by completing the
space of $C^2$ functions with respect to the norm $\|.\|$. With
the function $\xi$ at hand, we immediately solve the problem with
clamped plate we have started with. The deflection of the plate is
$w = \xi$; that is, the plate remains flat, only the point $r = 0$
is pulled out of it. The deformation energy is obtained most
easily from $U_d = 2\pi A$ by inserting $p = 0$ into the formula
for $A$. In this way we find
\begin{equation*} U_d = 2\pi \sqrt{2(1 - \kappa)}. \tag{18}
\end{equation*}

\vskip 2mm The search for the solution describing an infinite
plate leads to the conclusion that an infinite plate fixed on the
height 1 at $r = 0$ and on the height 0 at $r = 1$ has the
deflection $w = \xi_\infty \equiv \xi$ extended to all $r$. In
fig. 6, the infinite rigid-plastic plate lifted at the center is
depicted by the solid line
\begin{figure}[ht]
\centerline{\includegraphics[height=6cm]{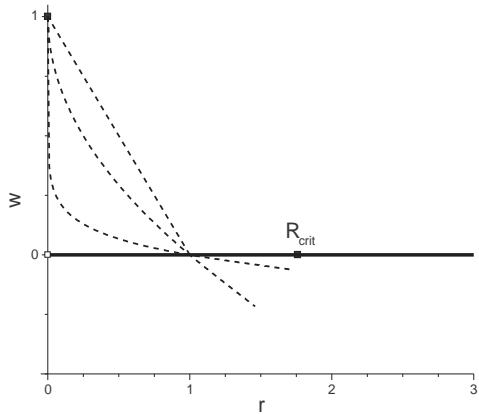}} \caption{An
infinite rigid-plastic plate}
\end{figure}
and by the black bullet on the vertical axis. To demonstrate the
transition to such plate, several finite rigid-plastic plates with
$\kappa = 0$ are shown in the figure, too, drawn by the dotted
lines. The curves were obtained by matching the solution $\thetag
{15}$ for $r \le 1$ with a quite intricate analytic solution for
$r > 1$. For the radius of the plate this procedure yields
$$R = \frac {\sqrt{1 + q}}{(1 + q^2)^{1/4}} \mbox
{exp} \left(\frac 12 \mbox {arctan} q\right).$$ If $p$ decreases
from 1 to 0, $q$ increases from 0 to 1 and $R$ increases from 1 to
$$R_{crit} = 2^{1/4} e^{\pi/8} \doteq 1.76.$$
The radius $R_{crit}$ is depicted in the figure by the bullet on
the horizontal axis. For $R \ge R_{crit}$ the deflection of the
plate is $w = \xi_R \equiv \xi$ extended to the radius $R$, so
that the plate is represented by the solid line cut at $r = R$ and
the bullet on the vertical axis. Finally, in fig. 7 several
infinite elastoplastic plates with $\kappa = 0$ are shown,
\begin{figure}[ht]
\centerline{\includegraphics[height=6cm]{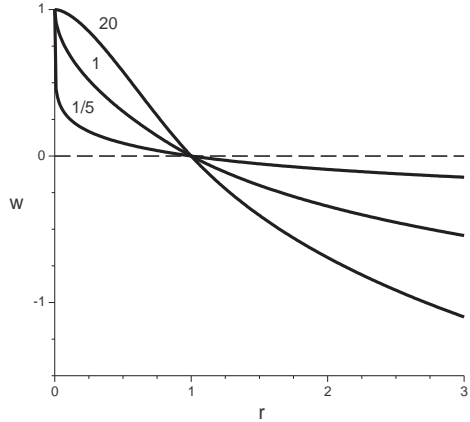}}
\caption{Infinite elastoplastic plates}
\end{figure}
with the values of the ratio ${\cal D}/D$ given next to them. The
solution can be expressed in terms of a set of parameters that are
fixed by a system of algebraic equations. Deformation energy is
$2\pi (\tilde q - \beta R_{plast}^2/2)$, where $\beta = {\cal
D}/D$ and $R_{plast}$ is the radius of the plastic domain in the
central part of the plate.

\vskip 2mm {\bf IX. Conclusion}

\vskip 2mm The mechanics of an elastic plate and the mechanics of
a rigid-plastic plate with the elliptical criterion look at first
glance similar: the latter differs from the former just by the
square root in the expression for the deformation energy. However,
this difference has far reaching consequences. Analyzing the
differential equation for deflection in the neighborhood of the
$\delta$-function source, one finds that the elastic plate
interpolates smoothly between the points at which it is fixed,
while the rigid-plastic plate has sharp vertices at these points.
Moreover, the solution for a circular rigid-plastic plate suggests
that if the size of the plate exceeds some limit value, the forces
fixing the plate at the given set of points fail to deform the
plate in the ordinary sense. The plate remains flat, and the
forces just pull the points out of it. Using the limit analysis it
can be shown that if the plate is infinite, it is not able to
reach equilibrium but in this peculiar way.

\vskip 2mm The behavior of the plastic plate can be also compared
to that of the plastic beam. Deformation energy of the plastic
beam is the total variation of deflection, which implies that the
beam fixed at the given set of points relaxes to a broken line. We
can see that if we pass from one dimension to two, the plastic
behavior becomes more singular.

\vskip 2mm {\it Acknowledgement.} This work was initiated 11 years
ago by Ivan Mizera, then professor of mathematical statistics at
Comenius University, who was interested in mechanical motivations
of some techniques in mathematical statistics. I am grateful to
him for many stimulating discussions a for his hospitality during
my one month stay at the University of Alberta. The stay was
funded by the grant VEGA 1/1008/09 and the NSERC of Canada.

\vskip 1cm {\bf References}

\vskip 4mm \noindent R. Franke (1985), Comp. Aid. Geom. Des. {\bf
2}, 87.

\vskip 2mm \noindent R. L. Harder and R. N. Desmarais (1972), J.
Aircraft {\bf 9}, 189.

\vskip 2mm \noindent R. Hill (1050): {\it The Mathematical Theory
of Plasticity}, Oxford University Press, Oxford.

\vskip 2mm \noindent L. M. Kachanov (1956): {\it Osnovy teorii
plastichnosti}, GITTL, Moscow; English translation: {\it Funda-
\linebreak \hglue 4mm mentals of the Theory of Plasticity},
North-Holland Publishing Company, Amsterdam (1971).

\vskip 2mm \noindent L. D. Landau and Y. M. Lifshitz (1965): {\it
Teoria uprugosti}, Moscow, Nauka; English translation: \linebreak
\hglue 4mm {\it Theory of Elasticity}, Pergamon Press, Oxford
(1975).

\vskip 2mm \noindent E. H. Mansfield (1957), Proc. Roy. Soc. {\bf
A 241}, 311.

\vskip 2mm \noindent L. Mit\'a\v s and H. Mit\'a\v sov\'a (1988),
Comp. Mat. App {\bf 16}, 983.

\vskip 2mm \noindent R. Temam (1983): {\it Probl\' emes
math\'ematiques en plasticit\'e,} Gautier-Villars, Paris.

\vskip 2mm \noindent S. P. Timoshenko and S. Woinowsky-Krieger
(1959): {\it Theory of plates and shells,} McGraw-Hill, New
\linebreak \hglue 4mm York.

\vskip 2mm \noindent W. H. Yang (1980a, b), J. App. Mech. {\bf
47}, 297, 301.

\end{document}